\documentclass[12pt]{article}
\usepackage{amssymb}
\usepackage{amsmath}

\begin{document}
\title{Supersymmetric Modified Korteweg-de Vries Equation: Bilinear Approach}
\author{Q.\ P.\  Liu$^\dagger$, Xing-Biao Hu$^\ddagger$ and Meng-Xia Zhang$^\dagger$\\
 \em\small $^\dagger$Department of Mathematics, \\
\em \small China University of Mining and Technology\\
\small\em Beijing 100083, P.R. China\\
\em\small and \\
$^\ddagger$\em\small Institute of Computational Mathematics, AMSS,\\
\em\small Academia Sinica, PO Box 2719, Beijing 100080,
P.R. China.\\
\em\small and\\
\em\small International Centre for Theoretical Physics, \\
\em\small Strada Costiera 11, 34014 Trieste, Italy}

\date{} \maketitle
\begin{abstract}
A proper bilinear form is proposed for the $N=1$ supersymmetric
modified Korteweg-de Vries equation. The bilinear B\"{a}cklund
transformation for this system is constructed. As applications,
some solutions are presented for it.
\end{abstract}

\section{Introduction}
The celebrated Korteweg-de Vries (KdV) equation is one of the most
important systems in mathematical physics. It has wide
applications and numerous interesting properties, such as soliton
solutions, infinite number of  conservation laws, B\"{a}cklund and
Darboux transformations, solvability in terms of inverse
scattering transformation. The KdV equation has many extensions
and one of them is the supersymmetric KdV equation constructed by
Manin and Radul \cite{mr} (see also \cite{mathieu}). Since then,
this $N=1$ supersymmetric KdV (sKdV) system has been studied
extensively and many interesting properties have been established.
For example, it is shown that sKdV equation has a bi-Hamiltonian
structure \cite{op}, Painlev\'{e} property\cite{ma}, infinitely
many symmetries, Darboux transformation \cite{liu} and
B\"{a}cklund transformation (BT) \cite{lx} and bilinear forms
\cite{my}\cite{ca}\cite{cgr}.

Closely related to the sKdV equation, the $N=1$ modified
Korteweg-de Vries (sMKdV) equation is introduced by Mathieu
\cite{mathieu} and Yamanaka and Sasaki \cite{sasaki})
respectively. It reads as
\begin{equation}
\label{smkdv1} \Psi_t+\Psi_{xxx}-3\Psi({\cal D}\Psi_x)({\cal
D}\Psi)-3({\cal D}\Psi)^2\Psi_x=0,
\end{equation}
where the field $\Psi=\Psi(x,\theta,t)$ is a Grassmann odd
variable depending on the space variables $(x, \theta)$ and time
$t$, ${\cal
D}={\partial\over\partial\theta}+\theta{\partial\over\partial x}$
is the usual super derivative. It is shown that equation
(\ref{smkdv1}) is related to the $N=1$ supersymmetric KdV through
a Miura type of transformation \cite{mathieu}\cite{sasaki}. The
sMKdV equation shares the common conserved quantities with the
supersymmetric sine-Gordon equation.

It is known that Hirota's bilinear approach is a very effective
method to construct particular solutions for soliton systems
\cite{hirotabk}. This method has been extended to supersymmetric
case in \cite{my}\cite{cgr}. In particular, Carstea, Grammaticos
and Ramani constructed the soliton type of solutions for the $N=1$
sKdV equation. In a recent paper \cite{gs}, the bilinearization of
the sMKdV equation is considered, namely the system (\ref{smkdv})
is transformed into
\begin{eqnarray*}
(SD_t+SD_x^3)(\tau_1\cdot\tau_1)&=&0,\\
(SD_t+SD_x^3)(\tau_2\cdot\tau_2)&=&0,\\
SD_x(\tau_1\cdot\tau_2)&=&0,\\
D_x^2(\tau_1\cdot\tau_2)&=&0,
\end{eqnarray*}
via $\Psi={\cal D}\ln{\tau_1\over\tau2}$, where the Hirota
derivative is defined as
\[
SD_t^mD_x^n f\cdot g=({\cal D}_{\theta_1}-{\cal
D}_{\theta_2})\left({\partial\over\partial
t_1}-{\partial\over\partial
t_2}\right)^m\left({\partial\over\partial
x_1}-{\partial\over\partial x_2}\right)^n
f(x_1,t_1,\theta_1)g(x_2,t_2,\theta_2)\left|_{\substack {x_1=x_2\\
t_1=t_2\\ \theta_1=\theta_2}}\right..
\]
Since there are only two tau functions $\tau_1$ and $\tau_2$, four
bilinear equations result in more restrictions than necessary.

The purpose of the present paper is to present our results on the
bilinear approach to the sMKdV equation. We will show that there
is indeed a proper bilinearization for this equation.

The paper is organized as follows. In the next section, we will
transform the sMKdV equation into bilinear form. In section 3, we
construct a bilinear B\"{a}ckulund transformation for the sMKdV
system. Interestingly, we will show that this B\"{a}ckulund
transformation in turn provides us a Lax operator for the sMKdV
equation. Then section 4 will be devoted to the construction of
some solutions. Final section contains our discussion and
conclusion.

\section{Bilinearzation}
It is noted that the sMKdV equation (\ref{smkdv1}) can be
rewritten as
\begin{equation}\label{smkdv}
\Psi_t+{\cal D}\Big[{\cal D}\Psi_{xx}+3\Psi\Psi_x({\cal
D}\Psi)-2({\cal D}\Psi)^3 \Big]=0,
\end{equation}
which suggests the following substitution
\[
\Psi={\cal D}\Phi,
\]
thus the system  (\ref{smkdv}) is transformed into its potential
form
\begin{equation}\label{psmkdv}
\Phi_t+\Phi_{xxx}-2\Phi_x^3+3({\cal D}\Phi)({\cal
D}\Phi_x)\Phi_x=0.
\end{equation}

To get its bilinearization, we take
\begin{equation}\label{sub}
\Phi=\ln{g\over f}
\end{equation}
where $f$ and $g$ are two Grassmann even functions. Then
straightforward calculation yields
\begin{equation}\label{sub1}
({\cal D}\Phi)({\cal D}\Phi_x)=\left[{{\cal D}(fg)\over
f^2g^2}\right]{\mathbb{B}}+{D_x^2(g\cdot f) -{\cal D}{\mathbb{B}}
\over fg},
\end{equation}
where
\begin{equation*}
\mathbb{B}=SD_x(g\cdot f),
\end{equation*}
while \begin{equation}\label{sub2}
 \Phi_t+\Phi_{xxx}-2\Phi_x^3=
{(D_t+D_x^3)(g\cdot f)\over fg}-3{(D_x^2g\cdot f)D_x(g\cdot
f)\over f^2g^2},
 \end{equation}
\begin{equation}\label{sub3}
 \Phi_x={D_x{g\cdot f}\over fg}.
\end{equation}

Substituting above expressions (\ref{sub1}-\ref{sub3}) into the
equation (\ref{psmkdv}), we obtain
\begin{eqnarray}
\Phi_t+\Phi_{xxx}-2\Phi_x^3 +3({\cal D}\Phi)({\cal D}\Phi_x)\Phi_x
&=& {(D_t+D_x^3)(g\cdot f)\over
fg}+\nonumber\\
&&+3\left[{{{\cal D}(fg)\over f^2g^2}}\mathbb{B}-{{\cal
D}\mathbb{B}\over fg}\right]{(D_xg\cdot f)\over fg}{},
\end{eqnarray}
Therefore, we have the following bilinearization for the sMKdV
equation
\begin{eqnarray}\label{bi1}
(D_t+D_x^3)(g\cdot f)&=&0,\\
\label{bi2} SD_x(g\cdot f)&=&0.
\end{eqnarray}

\noindent {\em Remark}: Unlike the previous attempt in \cite{gs},
our bilinearization (\ref{bi1}-\ref{bi2}) constitutes two
equations for two tau functions. This bilinearization is a natural
generalization of those for the classical MKdV equation
\cite{hirota}.

\section{B\"{a}cklund Transformation}
It is well known that BT is an useful concept and effective tool
for soliton systems as well as a characteristic of integrability.
In this section, we will derive a bilinear BT for our sMKdV
system. Our results are summarized in the following

\noindent {\bf Proposition} {\it Suppose that $(f, g)$ is a
solution of the equations (\ref{bi1}-\ref{bi2}), then $(\bar{f},
\bar{g})$ satisfying the following relations}
\begin{eqnarray}\label{bt1}
D_x{f}\cdot\bar{g}-\lambda D_xg\cdot\bar{f}=\mu f\bar{g}-\lambda\mu\bar{f}g,&&\\
\label{bt2} S{f}\cdot\bar{g}+\lambda Sg\cdot\bar{f}=\nu f\bar{g}+\lambda\nu\bar{f}g,&&\\
\label{bt3}
(D_t+D_x^3-3\mu D_x^2+3\mu^2 D_x)f\cdot\bar{f}=0,&&\\
\label{bt4} (D_t+D_x^3-3\mu D_x^2+3\mu^2 D_x)g\cdot\bar{g}=0,&&
 \end{eqnarray}
{\it is the another solution of (\ref{bi1}-\ref{bi2}), where
$\lambda$,
 $\mu$ are ordinary (even) constants and $\nu$ is an odd
constant}.

 {\em Proof}: We consider the following
\begin{eqnarray*}
 \mathbb{P}_1&=&2\Big[[SD_x f\cdot g]\bar{f}\bar{g}-fg[SD_x\bar{f}\cdot\bar{g}]\Big],\\
\mathbb{P}_2&=&[(D_t+D_{x}^3)f\cdot g]
 \bar{f}\bar{g}+fg[(D_t+D_{x}^3)\bar{g}\cdot\bar{f}].
\end{eqnarray*}
We will show that above equations (\ref{bt1}-\ref{bt4}) imply
$\mathbb{P}_1=0$ and $\mathbb{P}_2=0$. We first work on the case
of $\mathbb{P}_1$. We will use various bilinear identities which,
for convenience, are presented in the Appendix.
\begin{eqnarray*}
\mathbb{P}_1&\overset{(\ref{A1})}{=}&S\Big[(D_x f\cdot
\bar{g})\cdot \bar{f}g+f\bar{g}\cdot (D_x \bar{f}\cdot g)\Big]+
D_x\Big[(S f\cdot \bar{g})\cdot \bar{f}g+f\bar{g}\cdot (S \bar{f}\cdot g)\Big]\\
&\overset{(\ref{bt1}-\ref{bt2})}{=}&S\Big[(\lambda D_x g\cdot\bar{f}+\mu f\bar{g})\cdot\bar{f}g+f\bar{g}\cdot(-{1\over \lambda}D_x f\cdot\bar{g}-\mu \bar{f}g)\Big]+\\
&&+D_x\Big[(-\lambda Sg\cdot\bar{f}+\nu f\bar{g})\cdot\bar{f}g+f\bar{g}\cdot({1\over \lambda}Sf\cdot\bar{g}-\nu\bar{f}g)\Big]\\
&=&\lambda S[(D_xg\cdot\bar{f})\cdot\bar{f}g]-\lambda
D_x[(Sg\cdot\bar{f})\cdot\bar{f}g]-\\
&&-{1\over\lambda}S[f\bar{g}\cdot
(D_x{f}\cdot\bar{g})]+{1\over\lambda}D_x[f\bar{g}\cdot
(S{f}\cdot\bar{g})]\\
&\overset{(\ref{A2})}{=}&0.
\end{eqnarray*}

We now come to the second part of the proof.
\begin{eqnarray*}
\mathbb{P}_2&\overset{(\ref{A3}-\ref{A4})}{=}&(D_tf\cdot\bar{f})\bar{g}g+f\bar{f}(D_t\bar{g}\cdot
g)+(D_x^3f\cdot\bar{f})\bar{g}g+f\bar{f}(D_x^3\bar{g}\cdot
g)-\\
&&-3D_x[(D_x f\cdot\bar{g})\cdot (D_x g\cdot\bar{f})],
\end{eqnarray*}
but
\begin{eqnarray*}
D_x[(D_x f\cdot\bar{g})\cdot (D_x
g\cdot\bar{f})]&\overset{(\ref{bt1})}{=}& D_x[(\mu
f\bar{g}-\lambda\mu\bar{f}g)\cdot (D_xg\cdot\bar{f})]\\
&=&\mu D_x[f\bar{g}\cdot (D_x g\cdot\bar{f})]-\lambda\mu D_x[\bar{f}{g}\cdot (D_x g\cdot\bar{f})]\\
&\overset{(\ref{bt1})}{=}&\mu D_x[f\bar{g}\cdot (D_x
g\cdot\bar{f})]+\mu D_x[\bar{f}g\cdot(-D_x f\cdot\bar{g}+\mu
f\bar{g})]\\
&=&\mu D_x[(D_x f\cdot\bar{g})\cdot g\bar{f}+f\bar{g}\cdot (D_x
g\cdot\bar{f})]+\mu^2D_x(\bar{f}g\cdot f\bar{g})\\
&\overset{(\ref{A5})}{=}&\mu (D_x^2f\cdot\bar{f})g\bar{g}-\mu
f\bar{f}(D_x^2g\cdot\bar{g})+\mu^2D_x(\bar{f}g\cdot f\bar{g})\\
&\overset{(\ref{A6})}{=}&\mu (D_x^2f\cdot\bar{f})g\bar{g}-\mu
f\bar{f}(D_x^2g\cdot\bar{g})+\mu^2(D_x\bar{f}\cdot
f)g\bar{g}-\mu^2(D_x\bar{g}\cdot g)f\bar{f}\\
&=&[(\mu D_x^2-\mu^2 D_x)f\cdot\bar{f}]\bar{g}-f\bar{f}(\mu
D_x^2+\mu^2 D_x)\bar{g}\cdot{g},
\end{eqnarray*}
thus
\begin{equation*}
\mathbb{P}_2=[(D_t+D_x^3-3\mu D_x^2+3\mu^2
D_x)f\cdot\bar{f}]g\bar{g}-f\bar{f}[(D_t+D_x^3-3\mu D_x^2+3\mu^2
D_x)g\cdot\bar{g}]\overset{(\ref{bt3}-\ref{bt4})}{=}0.
\end{equation*}
this completes our proof.

\vspace{10pt}
 \noindent {\em Remark}: There are three constants in
our BT (\ref{bt1}-\ref{bt4}). In principle, these constants may
take arbitrary values, but to construct interesting solutions, we
are not allowed to set $\lambda$ to zero. Indeed, $\lambda$ can be
an arbitrary constant but zero and further may be renormalized to
unit. The true B\"{a}cklund parameter is $\mu$ as it will be clear
at the end of the section. \vspace{10pt}

Next we will demonstrate that a spectral problem can be derived
from above BT. To this end, we assume
\begin{equation}
u={\bar{f}\over f}, \;\;\; v={\bar{g}\over g},
\end{equation}
then by simple manipulation, we have
\begin{eqnarray}\label{sp1}
(v-\lambda u)_x&=&-\Phi_x(v+\lambda u)-\mu(v-\lambda u),\\
{\cal D}v+\lambda{\cal D}u&=&-({\cal D}\Phi)(v-\lambda
u)-\nu(v+\lambda u),
\end{eqnarray}
which constitute the spatial part of the spectral problem for our
system (\ref{smkdv}). To obtain a more compact form, we introduce
\[
U=v-\lambda u, \;\;\; V=v+\lambda u,
\]
in these variables, the spectral problem (\ref{sp1}) can be
rewritten simply as
\begin{eqnarray}\label{sp2}
U_x&=&-\Phi_x V-\mu U,\\
 {\cal D}V&=&-({\cal D}\Phi)U-\nu V.
\end{eqnarray}

It is interesting to note that we may have a scalar Lax operator
for the sMKdV equation. Indeed, letting $\nu=0$ and eliminating
one of the wave functions $V$, we arrive at the following Lax
operator
\begin{equation}\label{lax}
L=\partial_x-\Phi_x {\cal D}^{-1}({\cal D}\Phi),\;\; \mbox{or}\;\;
L=\partial_x-({\cal D}\Psi){\cal D}^{-1}\Psi,
\end{equation}
and now our system (\ref{smkdv}) has the Lax representation as
follows
\begin{equation}\label{lax2}
{d\over dt}L=[L,(L^3)_{\geq 0}].
\end{equation}
We remark here that the Lax operator (\ref{lax}) is in a form of
the constrained soliton systems. Indeed, this Lax operator is a
reduction of the so-called supersymmetric AKNS Lax operator
studied in \cite{aratyn}.

\section{Solutions}
For a given system, Hirota's bilinear form is ideal for
constructing particular solutions, so is B\"{a}cklund
transformation. For the sMKdV equation, we may adopt one of the
methods to find its soliton type of solutions. Since the
calculation involved here is straightforward although it is
cumbersome, we just list the results:

1-soliton:
\[
f=1+\exp{\eta},\;\; g=1-\exp{\eta},
\]
where $\eta=kx-k^3t+\theta\xi$ and $\xi$ is an arbitrary Grassmann
odd constant.

2-soliton:
\begin{eqnarray}
f&=&1+\exp(\eta_1)+\exp(\eta_2)+A_{12}\exp{(\eta_1+\eta_2}),\\
g&=&1-\exp(\eta_1)-\exp(\eta_2)+A_{12}\exp{(\eta_1+\eta_2}),
\end{eqnarray}
where
\begin{equation}
A_{12}=\left({k_1-k_2\over
k_1+k_2}\right)\Big[{k_1-k_2-2\xi_1\xi_2\over
k_1+k_2}+2\theta{(k_2\xi_1-k_1\xi_2)\over k_1+k_2}\Big],
\end{equation}
and
\[
\eta_i=k_ix-k_i^3t+\theta\xi_i,\; i=1,2
\]
and $\xi_1$, $\xi_2$ are arbitrary Grassmann odd constants. The
form of the last term can be reformed into the same one as
presented by Carstea, Ramani and Grammaticos for the
supersymmetric KdV equation \cite{cgr}.

\section{Conclusion and Discussion}
In this paper, we studied the sMKdV equation from the viewpoint of
Hirota's bilinear method. Contrast to the results of Ghosh and
Sarma \cite{gs}, we demonstrated that there exists a simpler
bilinearization for this system. We further obtained a bilinear
B\"{a}cklund transformation, which leads us to a new Lax operator
for the sMKdV equation.

We may convert the spatial part of our bilinear BT
(\ref{bt1}-\ref{bt2}) into the following form
\begin{equation}\label{bto}
\Phi_x+\bar{\Phi}_x+{\mu\over
2}[\exp{(\bar\Phi-\Phi)}-\exp{(\Phi-\bar\Phi)}]+({\cal
D}\Phi)({\cal D}\bar{\Phi}){\exp{\bar\Phi}-\exp{\Phi}\over
\exp{\bar\Phi}+\exp{\Phi}}=0,
\end{equation}
where we assume $\lambda=1$ and $\nu =0$. To compare with the BT
of MKdV equation, we let $\Phi=iQ$, $\bar{\Phi}=i\bar{Q}$, then
our BT (\ref{bto}) takes the following form
\begin{equation}
Q_x+\bar{Q}_x+\mu\sin{(\bar{Q}-Q)}-({\cal D}Q)({\cal
D}\bar{Q})\tan{\left({\bar{Q}-Q\over 2}\right)}=0,
\end{equation}
which is a generalization of the BT for the MKdV equation. It is
interesting to derive the corresponding nonlinear superposition
formula.

\vspace{10pt}
 {\bf Acknowledgements} The work is done when the
authors visited the Abdus Salam International Centre for
Theoretical Physics. We would like to thank the ICTP for support
and hospitality. QPL is supported in part by National Natural
Science Foundation of China under the grant number 10231050 and
the Ministry of Education of China, and XBH is supported by
National Natural Science Foundation of China under the grant
number 10171100.

\vspace{15pt}
\renewcommand{\theequation}{A.\arabic{equation}}
\setcounter{equation}{0}
\section*{Appendix: Some Bilinear Identities}
 In this Appendix, we list the relevant bilinear identities, which can be proved
 directly. Here $a$, $b$, $c$ and $d$ are arbitrary even functions
 of the independent variable $x$, $t$ and $\theta$.
\begin{eqnarray}\label{A1}
(SD_xa\cdot b)cd-ab(SD_xc\cdot d)&=&{1\over 2}S\Big[(D_x a\cdot
d)\cdot cb+ad\cdot (D_x c\cdot b)\Big]+\nonumber\\
&&+{1\over 2}D_x\Big[(S a\cdot d)\cdot cb+ad\cdot (S c\cdot
b)\Big] ,
\end{eqnarray}
\begin{equation}\label{A2}
S[(D_x a\cdot b)\cdot ab]=D_x[(S a\cdot b)\cdot ab],
\end{equation}
\begin{equation}\label{A3}
(D_x a\cdot b)cd+ab(D_x c\cdot d)=(D_x a\cdot d)cb+ad(D_x c\cdot
b),
\end{equation}
\begin{eqnarray}\label{A4}
(D_x^3a\cdot b)cd+ab(D_x^3c\cdot d)&=&(D_x^3 a\cdot d)\cdot
cb+ad\cdot (D_x^3 c\cdot b)-\nonumber\\
&&-3D_x(D_x a\cdot c)\cdot(D_x b\cdot d),
 \end{eqnarray}
\begin{equation}\label{A5}
D_x[(D_x a\cdot b)\cdot cd+ab\cdot(D_x c\cdot d)]=(D_x^2 a\cdot
d)cb-ad(D_x^2 c\cdot b),
\end{equation}
\begin{equation}\label{A6}
D_x ab\cdot cd=(D_x a\cdot d)cb-ad(D_x c\cdot b).
\end{equation}

\end{document}